\documentclass[11pt]{article}
\usepackage{blois,epsfig}

\def\Journal#1#2#3#4{{#1} {\bf #2}, #3 (#4)}

\def\PLB{{\em Phys. Lett.}  B}

\def\PRD{{\em Phys. Rev.} D}

\def\YAF{\em Yadernaya Fizika}
\def\GAC{\em Gravitation and Cosmology}
\def\GACS{{\em Gravitation and Cosmology} Suppl.}
\def\PAN{\em Phys. Atom. Nucl.}
\def\APP{\em Astropart. Phys.}
\def\JETP{\em JETP}
\def\NPPS{\em Nucl. Phys. Proc. Suppl.}

\def\be{\begin{equation}}
\def\ee{\end{equation}}
\def\bea{\begin{eqnarray}}
\def\eea{\end{eqnarray}}

\begin{document}
\vspace*{4cm}
\title{PHYSICS OF PRIMORDIAL UNIVERSE}

\author{ M. YU. KHLOPOV }

\address{Department of Physics, University "LaSapienza", Ple A.Moro,2,\\
Rome I-00185, Italy}

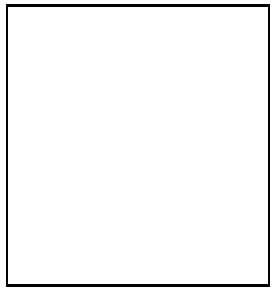
\maketitle\abstracts{ The physical basis of the modern
cosmological inflationary models with baryosynthesis and
nonbaryonic dark matter and energy implies such predictions of
particle theory, that, in turn, apply to cosmology for their test.
It makes physics of early Universe ambiguous and particle model
dependent. The study of modern cosmology is inevitably linked with
the probe for the new physics, underlying it. The particle model
dependent phenomena, such as unstable dark matter, primordial
black holes, strong primordial inhomogeneities, can play important
role in revealing the true physical cosmology. Such phenomena,
having serious physical grounds and leading to new nontrivial
cosmological scenarious, should be taken into account in the data
analysis of observational cosmology.}

\section{Cosmology of Particle Models}
In the modern cosmology the expansion of the Universe and its
initial conditions is related to the process of inflation. The
global properties of the Universe as well as the origin of its
large scale structure are the result of this process. The matter
content of the modern Universe is also originated from the
physical processes: the baryon density is the result of
baryosynthesis and the nonbaryonic dark matter represents the
relic species of physics of the hidden sector of particle theory.
Physics, underlying inflation, baryosynthesis and dark matter, is
referred to the extensions of the standard model, and the variety
of such extensions makes the whole picture in general ambiguous.
However, in the framework of each particular physical realization
of inflationary model with baryosynthesis and dark matter the
corresponding model dependent cosmological scenario can be
specified in all the details. In such scenario the main stages of
cosmological evolution, the structure and the physical content of
the Universe reflect the structure of the underlying physical
model. The latter should include with necessity the standard
model, describing the properties of baryonic matter, and its
extensions, responsible for inflation, baryosynthesis and dark
matter. In no case the cosmological impact of such extensions is
reduced to reproduction of these three phenomena only. The
nontrivial path of cosmological evolution, specific for each
particular realization of inflational model with baryosynthesis
and nonbaryonic dark matter, always contains some additional model
dependent cosmologically viable predictions, which can be
confronted with astrophysical data (see~\cite{book} for review).

\section{Cosmophenomenology of New Physics in Early Universe }

To study the imprints of new physics in astrophysical data the
forms and means in which new physics leaves such imprints should
be specified. So, the important tool in linking the cosmological
predictions of particle theory to observational data is the {\it
Cosmophenomenology} of new physics~\cite{Cosmology}. It studies
the possible hypothetical forms of new physics, which may appear
as cosmological consequences of particle theory, and their
properties, which can result in observable effects.

The simplest primordial form of new physics is the gas of new
stable massive particles, originated from early Universe. For
particles with the mass $m$, at high temperature $T>m$ the
equilibrium condition, $n \cdot \sigma v \cdot t > 1$ is valid, if
their annihilation cross section $\sigma > 1/(m m_{Pl})$ is
sufficiently large to establish the equilibrium. At $T<m$ such
particles go out of equilibrium and their relative concentration
freezes out. More weakly interacting species decouple from plasma
and radiation at $T>m$, when $n \cdot \sigma v \cdot t \sim 1$,
i.e. at $T_{dec} \sim (\sigma m_{Pl})^{-1}$. The maximal
temperature, which is reached in inflationary Universe, is the
reheating temperature, $T_{r}$, after inflation. So, the very
weakly interacting particles with the annihilation cross section
$\sigma < 1/(T_{r} m_{Pl})$, as well as very heavy particles with
the mass $m \gg T_{r}$ can not be in thermal equilibrium, and the
detailed mechanism of their production should be considered to
calculate their primordial abundance.

Decaying particles with the lifetime $\tau$, exceeding the age of
the Universe, $t_{U}$, $\tau > t_{U}$, can be treated as stable.
They form the modern multi-component dark matter, being its
dominant and sub-dominant components.

Primordial unstable particles with the lifetime, less than the age
of the Universe, $\tau < t_{U}$, can not survive to the present
time. But, if their lifetime is sufficiently large to satisfy the
condition $\tau \gg (m_{Pl}/m) \cdot (1/m)$, their existence in
early Universe can lead to direct or indirect traces. Cosmological
flux of decay products contributing into the cosmic and gamma ray
backgrounds represents the direct trace of unstable particles. If
the decay products do not survive to the present time their
interaction with matter and radiation can cause indirect trace in
the light element abundance or in the fluctuations of thermal
radiation. If the particle lifetime is much less than $1$s the
multi-step indirect traces are possible, provided that particles
dominate in the Universe before their decay. On the dust-like
stage of their dominance black hole formation takes place, and the
spectrum of such primordial black holes traces the particle
properties (mass, frozen concentration, lifetime) \cite{Polnarev}.
The particle decay in the end of dust like stage influences the
baryon asymmetry of the Universe. In any way cosmophenomenoLOGICAL
chains link the predicted properties of even unstable new
particles to the effects accessible in astronomical observations.
Such effects may be important in the analysis of the observational
data.

So, the only direct evidence for the accelerated expansion of the
modern Universe comes from the distant SN I data. The data on the
cosmic microwave background (CMB) radiation and large scale
structure (LSS) evolution (see e.g. \cite{WMAP}) prove in fact the
existence of homogeneously distributed dark energy and the slowing
down of LSS evolution at $z \leq 3$. Homogeneous negative pressure
medium ($\Lambda$-term or quintessence) leads to {\it relative}
slowing down of LSS evolution due to acceleration of cosmological
expansion. However, both homogeneous component of dark matter and
slowing down of LSS evolution naturally follow from the models of
Unstable Dark Matter (UDM) (see \cite{book} for review), in which
the structure is formed by unstable weakly interacting particles.
The weakly interacting decay products are distributed
homogeneously. The loss of the most part of dark matter after
decay slows down the LSS evolution. The dominantly invisible decay
products can contain small ionizing component \cite{Berezhiani2}.
Thus, UDM effects will deserve attention, even if the accelerated
expansion is proved.

The parameters of new stable and metastable particles are
determined by the pattern of particle symmetry breaking. This
pattern is reflected in the succession of phase transitions in the
early Universe. The phase transitions of the first order proceed
through the bubble nucleation, which can result in black hole
formation. The phase transitions of the second order can lead to
formation of topological defects, such as walls, string or
monopoles. The observational data put severe constraints on
magnetic monopole and cosmic wall production, as well as on the
parameters of cosmic strings. The succession of phase transitions
can change the structure of cosmological defects. The more
complicated forms, such as walls-surrounded-by-strings can appear.
Such structures can be unstable, but their existence can lead the
trace in the nonhomogeneous distribution of dark matter and in
large scale correlations in the nonhomogeneous dark matter
structures, such as {\it archioles} \cite{Sakharov2}.

\section{Strong Primordial Inhomogeneities }

The standard approach to LSS formation considers the evolution of
small initial fluctuations only. Such approach seems to be
supported by the homogeneity and isotropy of the Universe.
However, the amplitude of density fluctuations $\delta \equiv
\delta \varrho/\varrho$ measures the level of inhomogeneity
relative to the total density, $\varrho$. The partial amplitude
$\delta_{i} \equiv \delta \varrho_{i}/\varrho_{i}$ measures the
level of fluctuations within a particular component with density
$\varrho_{i}$, contributing into the total density $\varrho =
\sum_{i} \varrho_{i}$. The case $\delta_{i} \ge 1$ within the
considered $i$-th component corresponds to its strong
inhomogeneity. Strong inhomogeneity is compatible with the
smallness of total density fluctuations, if the contribution of
inhomogeneous component into the total density is small:
$\varrho_{i} \ll \varrho$, so that $\delta \ll 1$.

 The large scale correlations in topological defects and their imprints in
primordial inhomogeneities is the indirect effect of inflation, if
phase transitions take place after reheating of the Universe.
Inflation provides in this case the equal conditions of phase
transition, taking place in causally disconnected regions.

If the phase transitions take place on inflational stage new forms
of primordial large scale correlations appear. The example of
global U(1) symmetry, broken spontaneously in the period of
inflation and successively broken explicitly after reheating, was
recently considered in \cite{KRS}. In this model, spontaneous U(1)
symmetry breaking at inflational stage is induced by the vacuum
expectation value $\langle \psi \rangle = f$ of a complex scalar
field $\Psi = \psi \exp{(i \theta)}$, having also explicit
symmetry breaking term in its potential $V_{eb} = \Lambda^{4} (1 -
\cos{\theta})$. The latter is negligible in the period of
inflation, if $f \gg \Lambda$, so that there appears a valley
relative to values of phase in the field potential in this period.
Fluctuations of the phase $\theta$ along this valley, being of the
order of $\Delta \theta \sim H/(2\pi f)$ (here $H$ is the Hubble
parameter at inflational stage) change in the course of inflation
its initial value within the regions of smaller size. Owing to
such fluctuations, for the fixed value of $\theta_{60}$ in the
period of inflation with {\it e-folding} $N=60$ corresponding to
the part of the Universe within the modern cosmological horizon,
strong deviations from this value appear at smaller scales,
corresponding to later periods of inflation with $N < 60$. If
$\theta_{60} < \pi$, the fluctuations can move the value of
$\theta_{N}$ to $\theta_{N} > \pi$ in some regions of the
Universe. After reheating, when the Universe cools down to
temperature $T = \Lambda$ the phase transition to the true vacuum
states, corresponding to the minima of $V_{eb}$ takes place. For
$\theta_{N} < \pi$ the minimum of $V_{eb}$ is reached at
$\theta_{vac} = 0$, whereas in the regions with $\theta_{N} > \pi$
the true vacuum state corresponds to $\theta_{vac} = 2\pi$. For
$\theta_{60} < \pi$ in the bulk of the volume within the modern
cosmological horizon $\theta_{vac} = 0$. However, within this
volume there appear regions with $\theta_{vac} = 2\pi$. These
regions are surrounded by massive domain walls, formed at the
border between the two vacua. Since regions with $\theta_{vac} =
2\pi$ are confined, the domain walls are closed. After their size
equals the horizon, closed walls can collapse into black holes.
The minimal mass of such black hole is determined by the condition
that it's Schwarzschild radius, $r_{g} = 2 G M/c^{2}$ exceeds the
width of the wall, $l \sim f/\Lambda^{2}$, and it is given by
$M_{min} \sim f (m_{Pl}/\Lambda)^{2}$. The maximal mass is
determined by the mass of the wall, corresponding to the earliest
region $\theta_{N} > \pi$, appeared at inflational stage.  This
mechanism can lead to formation of primordial black holes of a
whatever large mass (up to the mass of AGNs ~\cite{AGN}). Such
black holes appear in the form of primordial black hole clusters,
exhibiting fractal distribution in space ~\cite{KRS}. It can shed
new light on the problem of galaxy formation.

Primordial strong inhomogeneities can also appear in the baryon
charge distribution. The appearance of antibaryon domains in the
baryon asymmetrical Universe, reflecting the inhomogeneity of
baryosynthesis, is the profound signature of such strong
inhomogeneity \cite{CKSZ}. On the example of the model of
spontaneous baryosynthesis (see \cite{Dolgov} for review) the
possibility for existence of antimatter domains, surviving to the
present time in inflationary Universe with inhomogeneous
baryosynthesis was revealed in \cite{KRS2}. Evolution of
sufficiently dense antimatter domains can lead to formation of
antimatter globular clusters \cite{GC}. The existence of such
cluster in the halo of our Galaxy should lead to the pollution of
the galactic halo by antiprotons. Their annihilation can reproduce
\cite{Golubkov} the observed galactic gamma background in the
range tens-hundreds MeV. The prediction of antihelium component of
cosmic rays~\cite{ANTIHE}, accessible to future searches for
cosmic ray antinuclei in PAMELA and AMS II experiments, as well as
of antimatter meteorites~\cite{ANTIME} provides the direct
experimental test for this hypothesis.

So the primordial strong inhomogeneities in the distribution of
total, dark matter and baryon density in the Universe is the new
important phenomenon of cosmological models, based on particle
models with hierarchy of symmetry breaking.

To conclude, the physical cosmology inevitably implies a set of
new nontrivial phenomena. Unstable dark matter, primordial black
holes and inhomogeneous structures, antimatter domains in baryon
asymmetrical Universe - are the examples of physically motivated
cosmological signatures of new physics, deserving attention in the
data analysis of precision cosmology.

\section*{Acknowledgments}
The work was performed in the framework of the State Contract
40.022.1.1.1106 and was partially supported by the RFBR grant
02-02-17490 and grant UR.02.01.026.

\end{document}